\documentclass{JINST}
\RequirePackage{xspace}
\def\epem {\ensuremath{e^+e^-}\xspace}
\def\emem {\ensuremath{e^-e^-}\xspace}
\def\gg {\ensuremath{\gamma \gamma}\xspace}
\def\ge {\ensuremath{\gamma e}\xspace}
\newcommand{\gev}{\ensuremath{\mathrm{\,Ge\kern -0.1em V}}\xspace}
\def\degc {\ensuremath{^\circ\mathrm{C}}\xspace}
\def\mus  {\ensuremath{\,\mu{\rm s}}\xspace}    

\title{A concept of the photon collider beam dump  }

\author{L.~I.~Shekhtman and V.~I.~Telnov\thanks{Corresponding author.} \\
Budker Institute of Nuclear Physics,\\
Novosibirsk State University, \\ 630090, Novosibirsk, Russia\\
E-mail: \email{telnov@inp.nsk.su}}

\abstract{Photon beams at photon colliders are very narrow, powerful (10--15 MW) and cannot be spread by fast magnets (because photons are neutral). No material can withstand such energy density. For the ILC-based photon collider, we suggest using a 150 m long, pressurized ($P\sim 4$ atm) argon gas target in front of a water absorber which solves the overheating and mechanical stress problems. The neutron background at the interaction point is estimated and additionally suppressed using a 20 m long hydrogen gas target in front of the argon.}


\begin{document}

\section{Introduction}\label{s1}
   Linear \epem colliders are generally considered to be the best tool for detailed study of new physics at energies $2E_0 \approx$ 0.2--3 TeV. They have been actively developed since 1980s. One of the projects, the ILC~\cite{ILC}, for the energy $2E_0=250$--500 GeV (and the potential to be upgraded to 1 TeV) is almost ready for construction in Japan and awaits approval. The second project, CLIC~\cite{CLIC}, for the energy up to 3 TeV, is also under development. The future of both projects depends on the energy scale of new physics. The Higgs boson and top quark give a rather good motivation for the ILC, while for the CLIC, additional motivation is needed. The next phase of LHC experiments, planned for 2015--2018, may help guide the decision.

   The photon collider (PC) based on conversion of electrons at linear colliders to high-energy photons using Compton scattering of laser photons has been discussed and actively developed since the early 1980s~\cite{GKST81,GKST83}. A PC would be a very natural and relatively inexpensive supplement to a high-energy \epem linear collider. The PC would enable the study of new physics in two additional types of collisions, \gg and \ge, at energies and luminosities close to those in \epem collisions.  A comprehensive description of the PC is given in the TESLA TDR~\cite{TESLA-PLC}; nearly everything regarding the PC at TESLA is also valid for the PC at the ILC, which we consider further.

\begin{figure}[!htb]
 \vspace{-0.cm}
\begin{center}
  \hspace{0.0cm}\includegraphics[width=10.5cm,height=9.5cm]{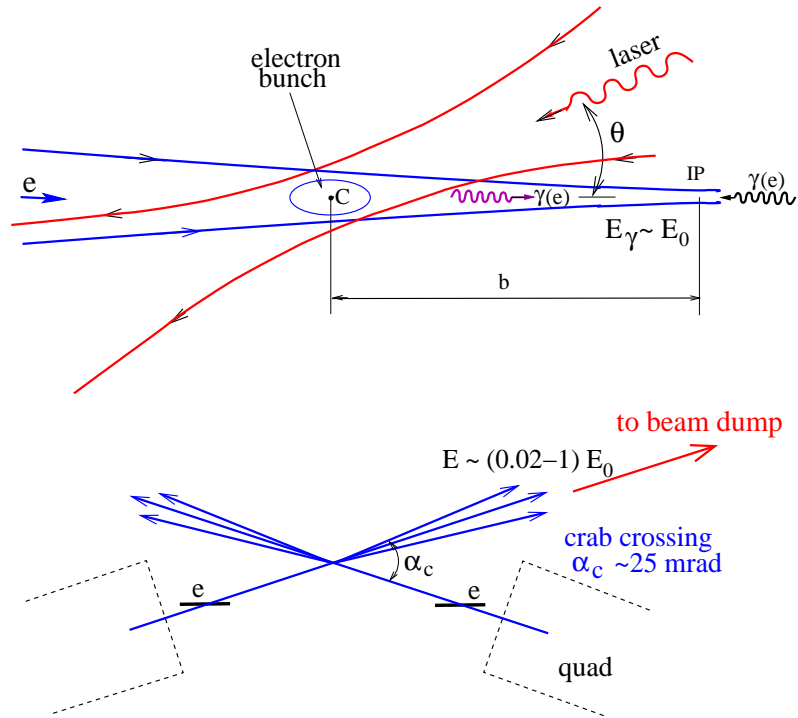}
\caption{\emph{(top)} The general scheme of a \gg, \ge photon collider. \emph{(below)} A crab-crossing collision scheme for the removal of disrupted beams from the detector to the beam dump. }
\end{center}
\label{scheme}
\vspace{-0.cm}
\end{figure}

   The general PC scheme is shown in Fig.~\ref{scheme} (top). Two electron beams, each of energy $E_0$, after passing the final focus system travel
towards the interaction point (IP). At a distance $b \sim \gamma \sigma_y$, or about 1--3 mm, from the IP, they collide with a focused laser beam.  After Compton scattering, the photons have an energy close to that of the initial electrons and follow their direction to the IP (with a small additional angular spread of the order of $1/\gamma$, where $\gamma =E_0/mc^2$), where they collide with a similar opposing beam of high-energy photons or electrons.  Using a picosecond laser with a flash energy of 5--10 joules, one can ``convert'' almost all electrons to high-energy photons. The photon spot size at the IP will be almost equal to that of the electrons, and therefore the total luminosity in \gg,
\ge collisions will be similar to the ``geometric'' luminosity of the underlying \epem beams (positrons are not necessary for photon colliders).

After crossing the conversion region, the electrons have a very broad energy spectrum, $E=($0.02--1)\,$E_0$, and large disruption
angles due to deflection of low-energy electrons in the field of the opposing beam. The removal of such a beam from the detector is therefore far
from trivial.

The ``crab crossing'' scheme of beam collisions solves the problem of beam removal at photon colliders~\cite{Telnov1990, Telnov1995},
Fig.~\ref{scheme} (bottom).  In the crab-crossing scheme~\cite{Palmer}, the beams are collided at a crossing angle $\alpha_\mathrm{c}$.  In order to preserve the
luminosity, the beams are tilted by a special RF cavity by the angle $\alpha_\mathrm{c}/2$.  If the crossing angle is larger than the disruption angles,
the beams just travel straight outside the quadrupole magnets.

After the IP, the disrupted beams have an angular spread of about $\pm 12 \mathrm{~mrad}$~\cite{TESLA-PLC,Telnov2006}. The disruption angle
for low-energy particles is proportional to $\sqrt{N/\sigma_z  E}$~~\cite{Telnov1990,Telnov1995} and depends very weakly on the transverse
beam size. The required crossing angle is determined by the disruption angle, the outer radius of the final quadrupole (about 5
cm~\cite{Telnov2006}), and the distance between the first quadrupole and the IP (about 4 m), which gives $\alpha_\mathrm{c} = 12 +5/400 \approx
25$ mrad.

Only one IP is planned in the present ILC design~\cite{ILC}, with a crossing angle of 14 mrad and two detectors in a pull-push configuration.
On the other hand, at the photon collider (PC) the beam-crossing angle must be at least 25 mrad. It seems quite reasonable to adjust the ILC design to make
the crossing angle $\sim 25$ mrad for both \epem and photon collisions in order to make it much easier and cheaper to transition from \epem to \gg or \ge modes of operation. Unfortunately, to minimize cost, the ILC team has developed only the minimal (baseline) \epem configuration, without the PC or any other ``options''. At present, we still have some time to change the beam crossing angle in the ILC design to make the ILC compatible with the future PC. This matter is currently under discussion.

   The present paper considers a narrower and more specific problem: how to dump used beams at photon colliders. It turned out that this problem is much more difficult than for \epem collisions.

The beam dumps at the \epem linear collider ILC ($2E_0 = 500-1000 \gev$) consist of water vessels 250 m from the interaction point~\cite{TESLA-beamdump,TESLA-TDR-acc}. The beam size at this distance is about $0.2 \times 0.7$ cm$^2$, while for the prevention of water boiling the beam should have a minimum transverse r.m.s.\ size of about 4 cm.  This problem is solved by using a magnetic sweeping system that distributes the 3000 electron bunches of a 1 ms long train around a circle of the radius $\sim 10$ cm (for $2E_0=1$ TeV) at the entrance of the water beam dump, thus reducing the energy density by two orders of magnitude.

    However, this scheme is not suitable for a photon collider (PC) since a photon beam is electrically neutral. Characteristic beam parameters at the ILC-based PC~\cite{Telnov2006} are as follows: $N=2\times 10^{10}$, the number of bunches in one train $n_\mathrm{b}=2820$, $\Delta t = 337$ ns, repetition rate $f = 5$ Hz, the angular divergence $\sigma_{\theta_x} \sim 3 \cdot 10^{-5}$, $\sigma_{\theta_y} \sim 10^{-5}$ rad (determined by $\beta$ functions at the interaction point), that is, 0.7 x 0.25 cm$^2$ at the distance of 250 m, and the beam power is about 12 MW at $2E_0=500 \gev$. No material can withstand such energy density. In addition, beams at photon colliders present a mixture of electrons and photons. The Compton-scattered photons have a very narrow angular distribution, ${\mathcal{O}}(10^{-5}$) rad, while the low-energy component of disrupted electron beams is rather broad, ${\mathcal{O}}(10^{-3}$--$10^{-2}$) rad. When the TESLA TDR was completed in 2001, a number of linear collider experts declared that a PC is impossible to build due to the unsolvable beam-dump problem. In reality, a much more serious problem for the PC was the removal of highly disrupted beams from the detector; as mentioned above, it was solved by the invention of the crab-crossing collision scheme. As for the beam dump problem, one of the workable solutions was suggested by us in 2003--2004 and reported at many linear collider workshops~\cite{S-T} since then. Below is a summary of our studies.

\section{Properties of beams at photon colliders after the IP}\label{s2}
\begin{figure}[!htb]
\begin{minipage}{0.5\linewidth}
 \vspace{-0.8cm}
  \hspace{-0.5cm}\includegraphics[width=8.5cm,height=8.cm]{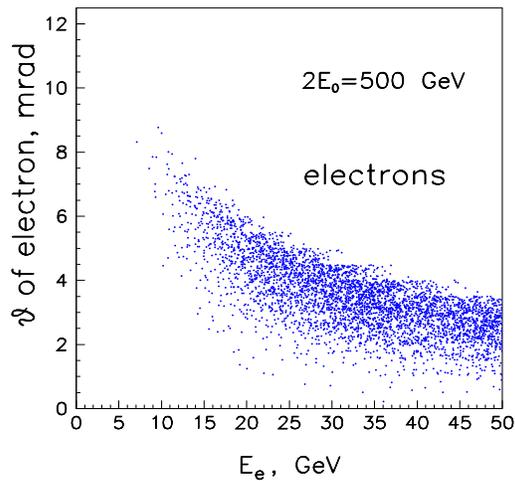}
\end{minipage} \hspace{0.5cm}
\begin{minipage}{0.43\linewidth} \hspace{0mm}
\caption{Angles of disrupted electrons after Compton
    scattering and interaction with the opposing electron beam; $N=2\times
    10^{10}$, $\sigma_z=0.3$ mm.}  \hspace{0.5cm}
\label{e-angle}
\end{minipage}
\vspace{-1.cm}
\end{figure}

\begin{figure}[!htb]
\begin{center}
 \hspace{-0.7cm}\includegraphics[width=8.2cm]{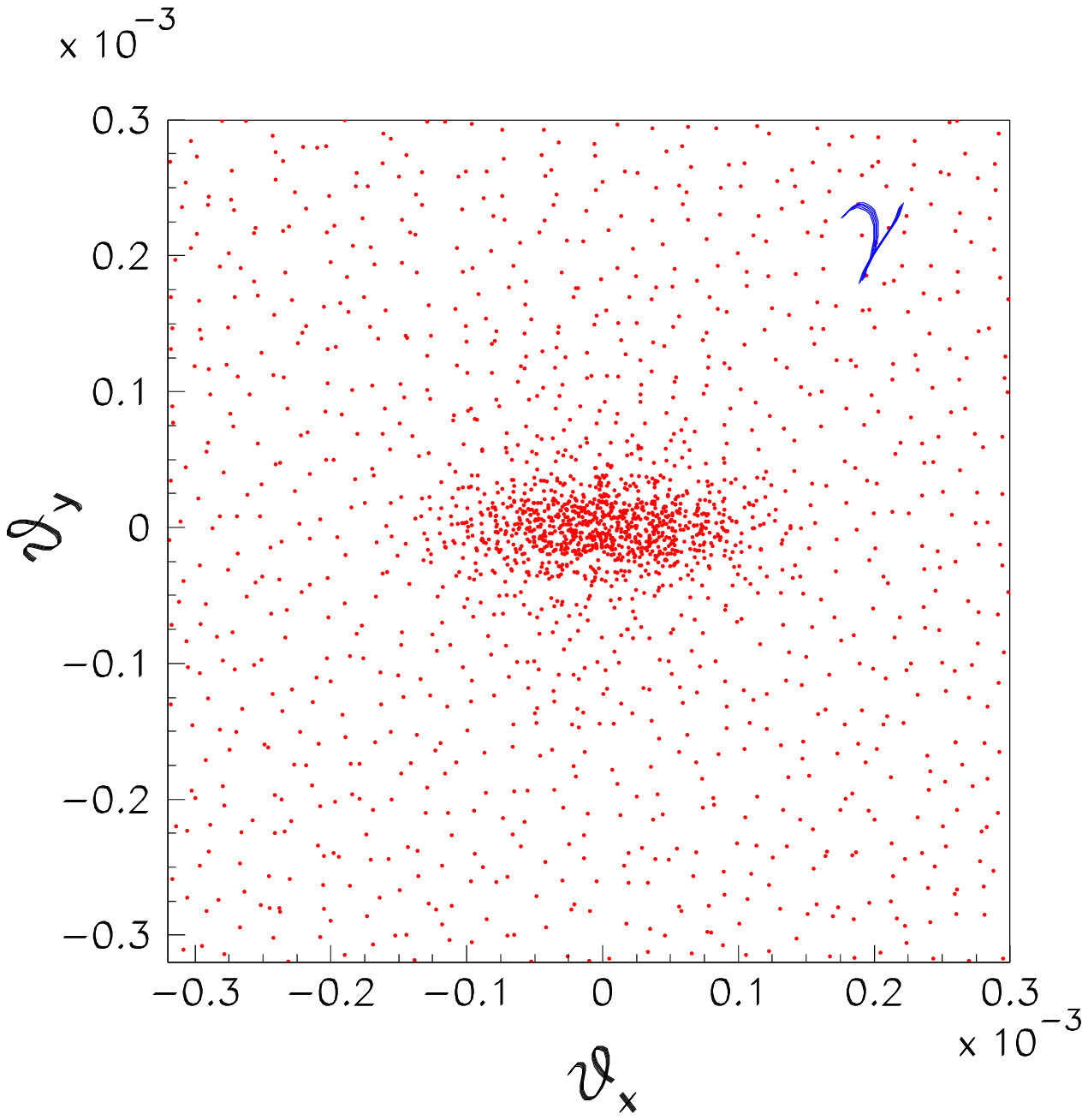}\hspace{-0.7cm}
\includegraphics[width=8.2cm]{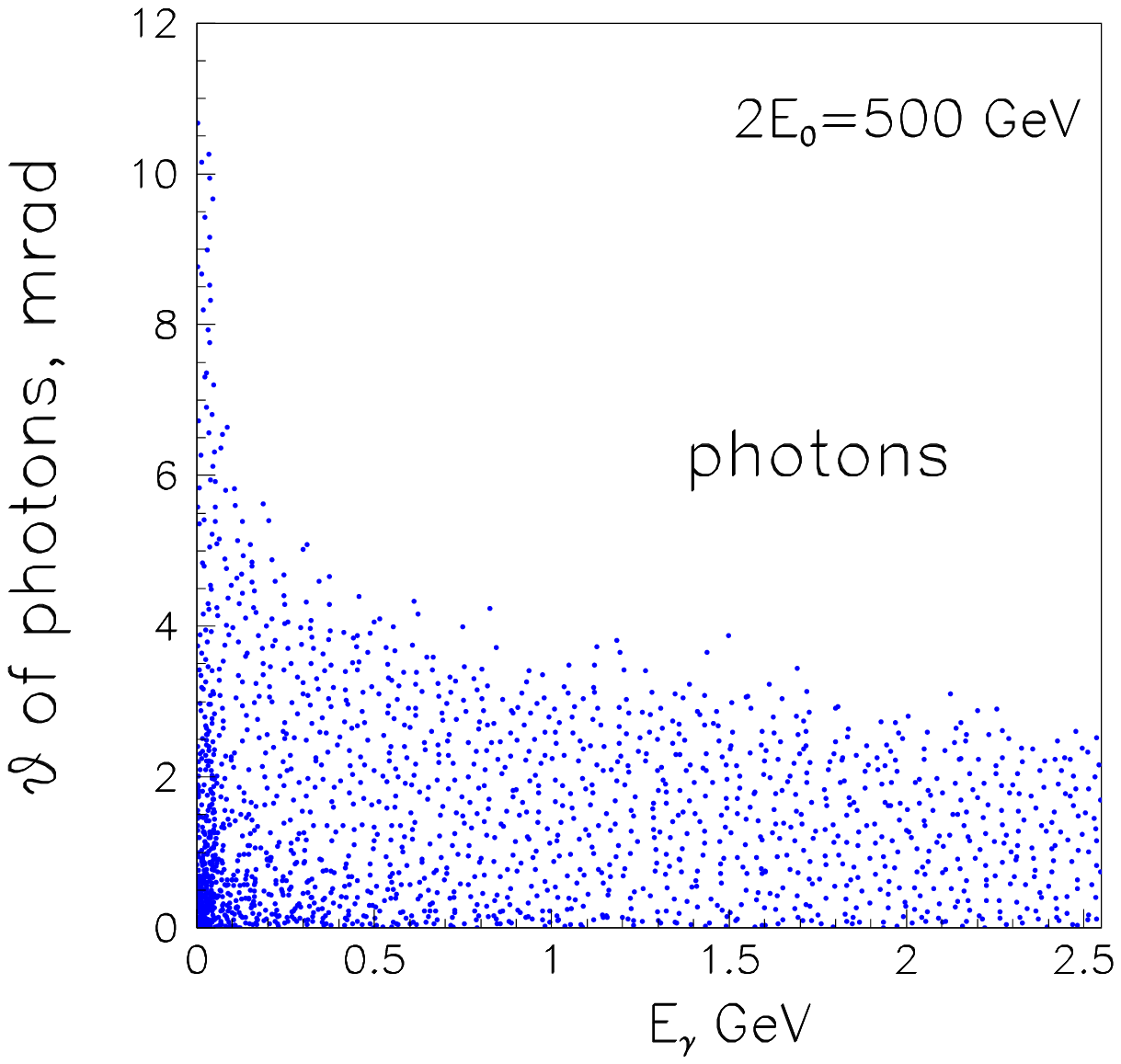} \hspace{-0.cm}
\end{center}
\vspace{-1.2cm}
\caption{\emph{(left)} The angular distribution of Compton photons; \emph{(right)} the energy-angular distribution of beamstrahlung photons.}
\label{p-angle}
\end{figure}

\begin{figure}[!htb]
\vspace{-1.cm}
\begin{center}
 \hspace{0cm} \includegraphics[width=7.7cm]{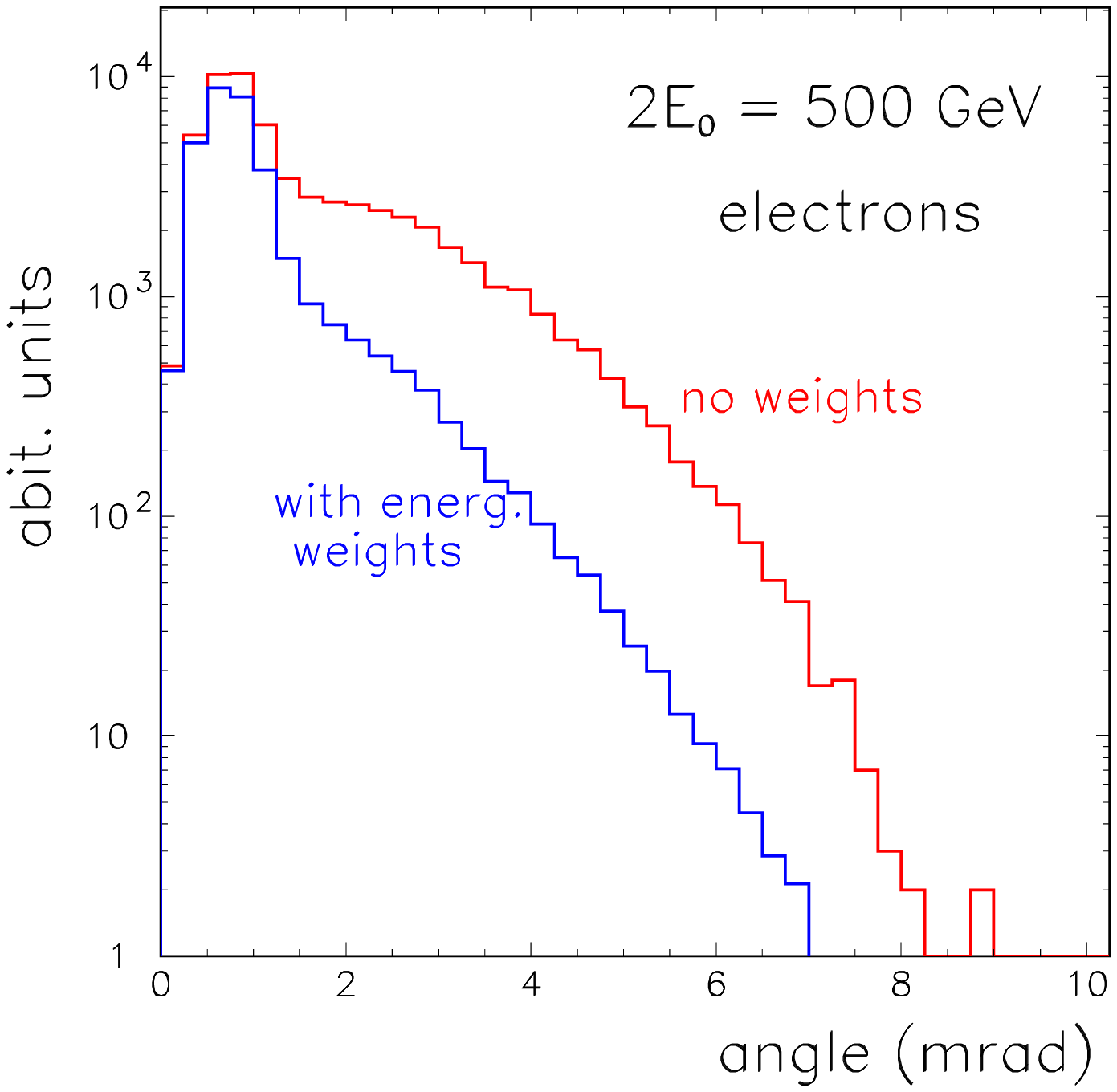} \hspace{-0.7cm}
\hspace{-0.cm}\includegraphics[width=7.7cm]{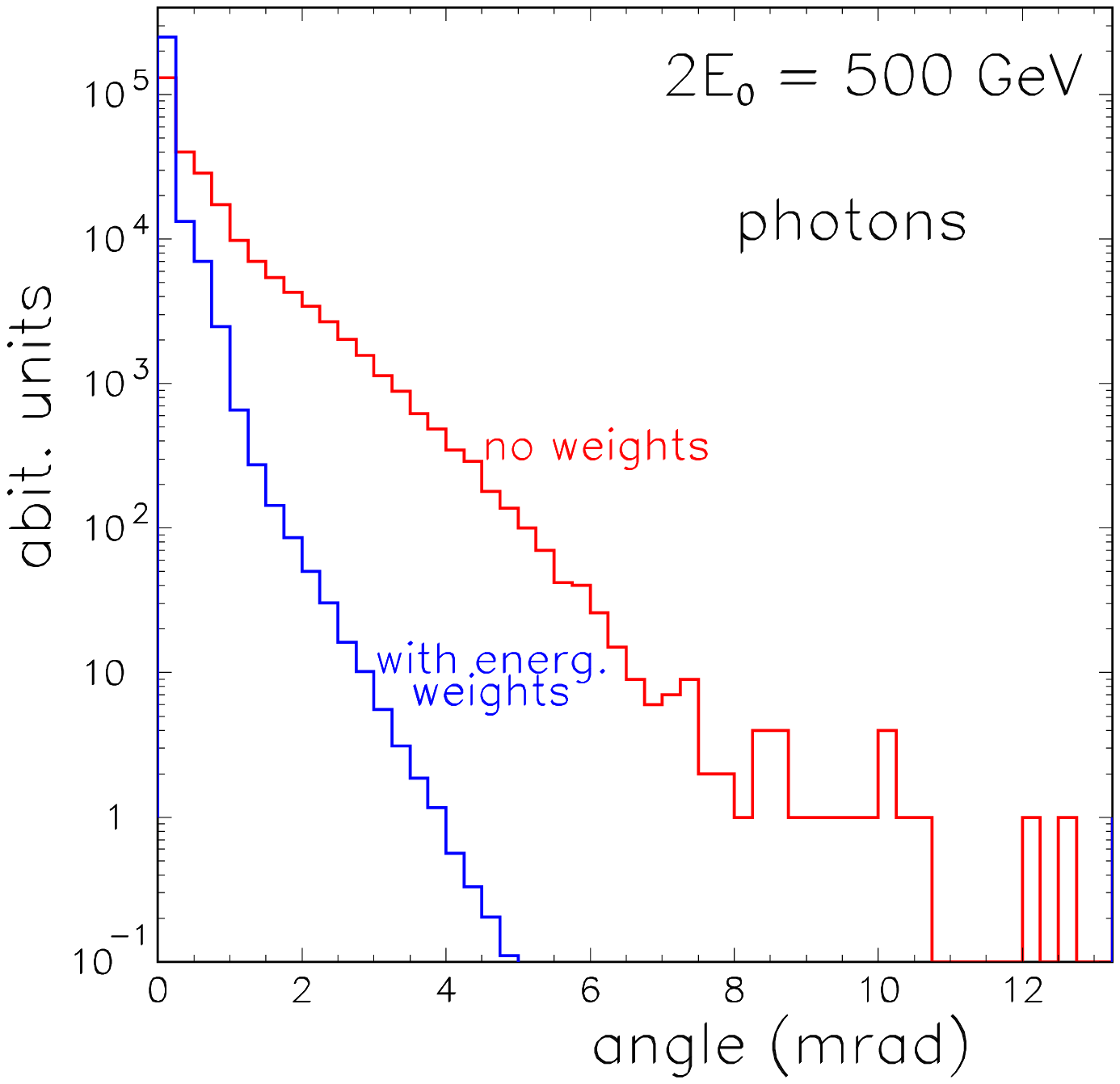}
\end{center}
\vspace{-0.5cm}
\caption{Angular distributions of electrons (left) and photons (right)
after the conversion and interaction points.}
\label{ang-dis}\vspace{-0.0cm}
\end{figure}

  Properties of outgoing beams at photon colliders, presented below, were obtained by our (V.I.T.) code described (briiefly) in Refs.~\cite{Telnov1995} and \cite{TESLA-PLC}. The code is capable of simulation of \epem, \emem, \ge, \gg\/  beam collisions in linear colliders and was used for photon-collider studies at NLC, TESLA, CLIC and ILC. It takes into account all important processes, including: \\[-5mm]
\begin{enumerate}
\item Compton scattering in the conversion region, including nonlinear and polarization effects.\\[-5mm]
\item \epem pair creation in the conversion region for $x>4.8$.\\[-5mm]
\item  Deflection by  magnetic fields and synchrotron radiation in the detector solenoidal field (due to the crab-crossing angle).\\[-5mm]
\item   Deflection of particles in the field of oncoming beams, coherent pair creation and beamstrahlung during beam collisions at the IP.\\[-5mm]
\end{enumerate}
The outgoing beam after the conversion (CP) and interaction points (IP) is a mixture of photons and electrons (plus some amount of positrons). About half of the energy is carried by electrons, the other half by photons. After Compton scattering, the electrons have a very wide energy spread, $E\sim$ (0.02--1)$E_0$. During the beam collision, these electrons are deflected by the opposing beam by angles of about 2--10 mrad, depending on their energy, Fig.~\ref{e-angle}~\cite{Telnov2006}.  The high-energy photons, obtained by Compton scattering of laser photons on high-energy electrons, have a very narrow angular distribution, about 0.01--0.03 mrad, Fig.~\ref{p-angle}~(left).  Electrons emit many beamstrahlung photons along their directions, Fig.~\ref{p-angle}~(right). Angular distributions and energy flows for electrons and photons are shown in Fig.~\ref{ang-dis}.
If the beam dump is situated at $L=250$ m, then for particles with $\theta=5$ mrad the beam radius at the beam-dump entrance is $r \sim 1.25$ m. Some focusing of electrons will be useful in order to reduce the diameter of the beam-dump tube. For photons, a clear angle of about 3 mrad will be sufficient, that is, $r\sim 75$ cm at $L=250$ m.  So, the outgoing beam has two components: very narrow ($0.25 \times 0.75$ cm$^2$) and very wide (up to $r \sim 1$ m) at the entrance to the beam dump, which causes additional complications.

\section{Beam dump scheme}\label{s3}
\begin{figure}[ht]
     \begin{center}
     \vspace*{-.cm}
     \includegraphics[width=13cm]{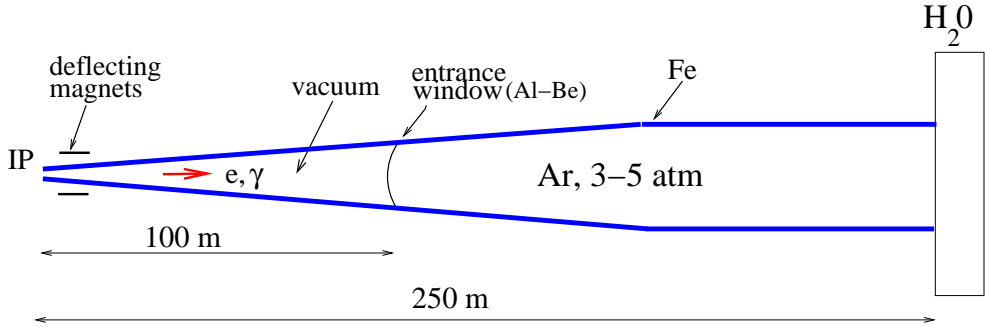}
     \end{center}
     \caption{An idea of the beam dump for the photon collider at ILC.}
     \label{scheme1}
     \end{figure}

\begin{figure}[!htb]
\centering
\includegraphics[width=13cm]{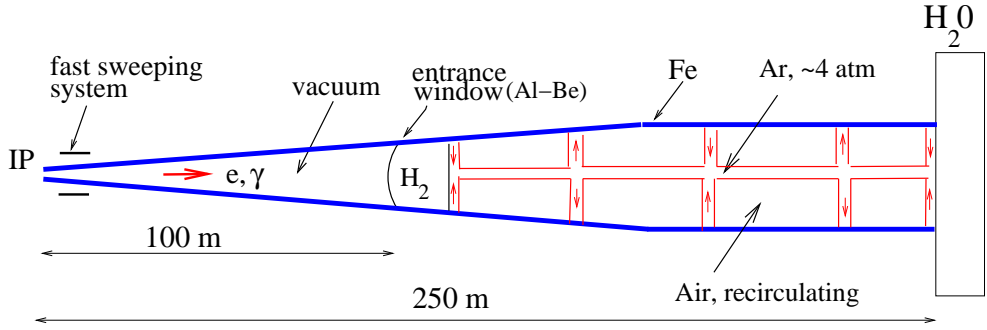}
\vspace{0.0cm}
\caption{The beam dump with a neutron absorber and a reduced-diameter Ar tube (see text).  }
\vspace{0.3cm}
\label{scheme2}
\end{figure}

 We propose a scheme of the beam dump for the photon collider at the ILC as depicted in Figs.~\ref{scheme1} and \ref{scheme2}.  In addition to the TESLA--ILC solution for \epem beams, we have added a gas beam target (converter, expander) of 4-5 radiation lengths thickness. A reasonable choice is a one atomic gas (to avoid radiolysis) argon at 3--5 atm ($X_0=110$ m at $P = 1$ atm). 

 The sweeping magnets are needed to decrease stress and heating in the thin Al--Be entrance window situated 100 m from the IP. A circle radius of $R= $0.5--1 cm at this point is sufficient for the hard charged component of the beam. As for the narrow photon component, the probability of \epem conversion in the entrance window is small and the energy deposition by produced charged particles is acceptable (see below). In the gas, photons and electrons produce showers. Due to multiple scattering, the density of particles at the exit window of the gas vessel and the entrance window of the water beam dump is also acceptable. The third critical place is the densest point of the shower situated in the water beam dump between the entrance window and the shower maximum. In the proposed scheme, the rise of the water temperature at this point is also acceptable.

   In principle, the diameter of the Ar target of about 10 cm is sufficient, and a shower of such diameter does not present a problem for the water beam dump.  The rest of the exit pipe volume with a diameter of about 1.5 m can be empty as shown in Fig.~\ref{scheme2}. The advantage of this scheme compared to the one in Fig.~\ref{scheme1} is the fact that electrons (especially low-energy electrons) traveling at large distances from the axis would not get as much additional scattering, which would avoid the unnecessary widening of the beam and reduce the leakage of radiation to unshielded areas.

\section{Heat loads}\label{s4}
\begin{figure}[ht]
     \begin{center}
     \vspace*{0.2cm}
\includegraphics[width=13.cm,angle=0]{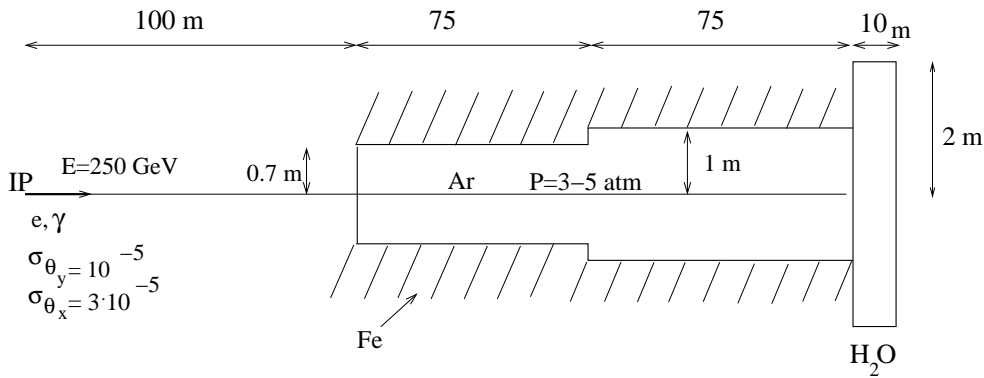}
    \vspace*{-0.3cm}
     \end{center}
     \caption{The scheme used in the simulation.}
   \vspace*{-0.cm}
   \label{scheme-simul}
   \end{figure}

The simulation using FLUKA code~\cite{FLUKA} was done for the geometry shown in Fig.~\ref{scheme-simul}.
  The results are as follows.  After the passage of a 1 ms bunch train consisting of 3000 bunches, the maximum temperature rise
   $\Delta T$ of the  Be--Al entrance window is about $40\degc$ for
   a sweeping radius $R=0.5$ cm. Thermal conductivity would be sufficient for the removal of the heat; gas cooling can be added if necessary.  The maximum local $\Delta T$ at the exit Be--Al window is small, about 10\degc.  The maximum local $\Delta T$ in the water dump from 250 GeV photons is 75,  50, 25 \degc at Ar pressures of 3, 4, 5 atm, respectively, and a factor of 2 lower from incident electrons. 
   
   In the case of the energy the upgrade from 250 to 500 GeV, the temperature of the entrance window increases by a factor of 2 due to smaller photon spot size (for fixed $\beta$ functions at the IP); the increase of temperatures at the exit window and inside the water dump will be less than a factor of 2 (proportional to the shower density). Such increase of temperatures is acceptable, higher beam energy needs somewhat higher Ar pressure.
   
The problem of mechanical stress in solid materials in the TESLA beam dump is not important because the train duration (1 ms) is much longer
than the decay time of local stress ($r/v_\mathrm{sound} \sim 1 \mus$). It is more serious for ``warm'' LCs with short trains.

\section{Neutron background}\label{s5}
In our study we paid a special attention to the reduction of the neutron background. Neutrons produced in showers can travel in backward direction and damage the vertex detector. For the photon collider (PC), this problem is more serious than for \epem colliders because the PC has a large exit aperture for disrupted beams.

  Simulation with FLUKA gives an estimate of the neutron flux at the IP (it is a highly time-consuming simulation).  For $10^5$ incident 250 GeV electrons and P(Ar) $=4$ atm, on average only 6 neutrons cross the IP plane, $z=0$, with the radial coordinates $r=$ 1.5,
2.5, 4.5, 14.5, 18.5 21.5 m.  Due to the collimation by the Fe tube we do not expect the uniform density, the density per cm$^2$
should be larger near the axis. Assuming the uniform density for three
neutrons closest to the axis we find the flux of $\sim 5 \times 10^{-11}$ $n$/cm$^2$ per incident electron, or about $1.5 \times 10^{11}$ $n$/cm$^2$ for 10$^7$ s run time.

   It is remarkable that after the replacement of the first 20 m of Ar by  H$_2$ at the same pressure, only one neutron crosses the IP plain at $r=1.5$ m per $8 \cdot 10^5$ incident electrons. With account of collimation  by the tube it means the decrease of the neutron flux at least by a
   factor of ten!
    
\section{Conclusion}
We have proposed and considered at the conceptual level a beam dump scheme that can withstand the very powerful and narrow beams at photon colliders. It consists of a water beam dump after a 150 m long Ar gas target at about 4 atm pressure that convert photons to \epem pairs and expands the beam diameter. This approach can also be used at high-energy \epem colliders where beamstralung photons carry a substantial part of the original beam energy. Adding a hydrogen-filled segment at the beginning of the beam-dump system very effectively protects the detector from the neutron background coming from the beam dump. If desired, this system could be equipped with some simple beam-profile diagnostics based, for example, on gas scintillations.

\section*{Acknowledgments}

The work was supported by the Ministry of Education and Science of the Russian Federation.

\end{document}